\documentclass[10pt]{article}
\usepackage[top=0.85in,left=0.75in,footskip=0.75in,marginparwidth=2in]{geometry}

\usepackage[T2A,T1]{fontenc}
\usepackage[utf8]{inputenc}
\usepackage[russian,french,english]{babel}

\usepackage{amsfonts}
\usepackage{amsmath,amssymb,amsfonts}%
\usepackage{amsthm}%
\usepackage{mathrsfs}%

\usepackage{nicefrac}
\usepackage{graphicx}
\usepackage{subfig}
\usepackage{hyperref}
\usepackage{url}            % simple URL typesetting
\usepackage{booktabs}  
\usepackage{nicefrac}
\usepackage{microtype}

\usepackage{graphicx}
\usepackage{subfig}
\newtheorem{definition}{Definition}
\graphicspath{ {./images/} }

\usepackage{cite}

\usepackage{nameref,hyperref}

\usepackage[right]{lineno}

\usepackage{microtype}
\DisableLigatures[f]{encoding = *, family = * }

\usepackage{setspace} 
\usepackage{changepage}

\usepackage[aboveskip=1pt,labelfont=bf,labelsep=period,singlelinecheck=off]{caption}

\makeatletter
\renewcommand{\@biblabel}[1]{\quad#1.}
\makeatother

\usepackage{lastpage,fancyhdr,graphicx}
\usepackage{epstopdf}
\fancyheadoffset[L]{2.25in}
\fancyfootoffset[L]{2.25in}

\usepackage{color}

\definecolor{Gray}{gray}{.25}

\usepackage{graphicx}

\usepackage{sidecap}

\usepackage{wrapfig}
\usepackage[pscoord]{eso-pic}
\usepackage[fulladjust]{marginnote}
\reversemarginpar

\begin{document}
\vspace*{0.35in}

% title goes here:
\begin{flushleft}
{\Large
\textbf\newline{$\Delta$-Conformity: Multi-scale Node Assortativity in Feature-rich Stream Graphs}
}

%\newline
% authors go here:
%\\
Salvatore Citraro \textsuperscript{1, 2},
Letizia Milli \textsuperscript{2},
Rémy Cazabet \textsuperscript{3},
Giulio Rossetti \textsuperscript{2},

%\\
\bigskip
\bf{1} Department of Computer Science, University of Pisa
  Largo Bruno Pontecorvo, 3, Pisa
\\
\bf{2} KDD-Lab, ISTI (CNR) G. Moruzzi, 1, Pisa
\\
\bf{3} University of Lyon 1, CNRS, LIRIS UMR 5205, F-69622 France
\\ 
\bigskip
* salvatore.citraro@phd.unipi.it

\end{flushleft}

\section*{Abstract}

Heterogeneity is a key aspect of complex networks, often emerging by looking at the distribution of node properties, from the milestone observations on the degree to the recent developments in mixing pattern estimation.
Mixing patterns, in particular, refer to nodes' connectivity preferences with respect to an attribute label.
Social networks are mostly characterized by assortative/homophilic behaviour, where nodes are more likely to be connected with similar ones. Recently, assortative mixing is increasingly measured in a multi-scale fashion to overcome well-known limitations of classic scores.
Such multi-scale strategies can capture heterogeneous behaviors among node homophily, but they ignore an important, often available, \textit{addendum} in real-world systems: the time when edges are present and the time-varying paths they form accordingly.
Hence, temporal homophily is still little understood in complex networks.
In this work we aim to cover this gap by introducing the \textit{$\Delta$-Conformity} measure, a multi-scale, path-aware, node homophily estimator within the new framework of feature-rich stream graphs.
A rich experimental section analyzes \textit{$\Delta$-Conformity} trends over time, spanning the analysis from real-life social interaction networks to a specific case-study about the Bitcoin Transaction Network. 

% now start line numbers
%\linenumbers

\section{Introduction}\label{sec:intro}

Networks help scientists to represent phenomena of uncountable complexity. %, from biological systems to all human-specific activities related to our social instinct and complex cognition.
The \textit{network model} -- in its minimal definition of a set of nodes with edges linking them -- is acknowledged by researchers from many domains, leading to the consolidation of a multidisciplinary field crossing over physics, math, economy, and social sciences.
While analyzing complex data, the network topology is one of the most informative characteristics to focus on, making network science a solid paradigm for unveiling universal principles from different and domain-specific systems.
Nevertheless, a paradigm evolves as its research questions evolve: nowadays, increasingly new approaches aim to combine the expressive power of topology to all those domain-specific aspects often available from complex systems.
From attributes describing nodes to the insightful addendum of the temporal dimension, such enriched aspects can enhance valuable knowledge hidden in complex systems.
The augmented structures incorporating them are often referred to as \textit{feature-rich networks} \cite{interdonato2019feature}. %, an umbrella-term generalizing several classes of network extensions, all of them unified by the aim to add more information to the graph topology.
In this work we focus on specific instances of such augmented topologies: \emph{node-attributed} and \emph{dynamic} graphs. 
The former aspect focuses on the study of networks whose nodes are semantically enriched by context dependent attributes, the latter on frameworks designed to analyze the evolution of complex systems over time.
%Introducing such a general term can aid researchers to experience more complex modelling environments and provide new methodologies to deal with them.

Leveraging such enriched network models we study the evolution of \textit{homophilic/heterophilic} behaviours - a critical emerging behavior in social systems - over time. 
Indeed, network science literature has deeply investigated homophily by introducing several measures to quantify it in networks, but little attention was given to its relation with the temporal dimension - i.e., studying to what extent it remains stable/changes as the underlying topology changes.
Being interested in tracking nodes' homophily over time, we leverage the combined expressive power of node-attributed and dynamic graphs -- these latter ones defined through the formalism of stream graphs \cite{latapy2018stream}.
In this work, we propose a node-centric, path-aware, homophily measure for attributed stream graphs, i.e., \textit{$\Delta$-Conformity}, as a conservative extension of \textit{Conformity} \cite{rossetti2021conformity}, a recent multi-scale measure aiming to capture the heterogeneity of mixing patterns in networks.
\textit{$\Delta$-Conformity} leverages the concept of time-respecting paths, and it is able to cope with both static and varying categorical attributes.
%, e.g., in a social network, an individual first language (static attribute) or its political leaning (varying attribute).

The rest of the paper is organized as follows.
Section \ref{sec:related} introduces the main literature needed to understand the current work fully;
Section \ref{sec:delta_conf} introduces and formally describes \textit{$\Delta$-Conformity};
Section \ref{sec:exp} provides several case studies on which the proposed formula can be applied;
Section \ref{sec:conc} concludes the work.

\section{Related Works} \label{sec:related}

An overview of several topics is provided here, namely i) homophily estimation in node-attributed static graphs, ii) dynamic networks representations, and iii) homophily definitions in dynamic environments.

\noindent {\bf Homophily estimation}. In social networks \textit{homophily} refers to the tendency of people to be more likely to interact with similar others w.r.t. several social dimensions, from age to political leaning, often grounded by households, workplaces, geographical environments. \cite{mcpherson2001birds}.
Studying homophilic behaviour may help researchers to unveil critical networks dynamics, as in the studies of i) segregation in interracial friendships \cite{moody2001race}, ii) gender-specific patterns in early school grades \cite{shrum1988friendship}, iii) sexually transmitted diseases spreading \cite{aral1999sexual}, iv) trustworthiness in business networks \cite{barone2018birds}.

The Newman's \textit{assortativity coefficient} $r$ \cite{newman2003mixing} is a popular measure that quantifies homophily in complex networks.
This quantity is calculated as the sum of the differences between the observed and the expected fraction of edges between nodes sharing similar values of an attribute.
When maximized, $r=1$, the assortativity coefficient describes a network where all edges connect to nodes labeled with the same value; when $r=0$, the edges are randomly connected; when minimized, $r=-1$, the coefficient describes a network where all edges connect to nodes with a different value.
Other assortativity measures, like \textit{ProNe} \cite{rabbany2017beyond}, or statistical approaches, like the \textit{VA-Index} \cite{pelechrinis2016va}, include more complex scenarios, estimating the correlation between structure and two or more attributes.

Nevertheless, a limitation of these measures is that they sum up a single, average network behavior, ignoring/absorbing the plausible existence of outliers and heterogeneous mixing patterns.
Considering individual connectivity preferences can shed light on important network dynamics, as in the studies on i) \textit{monophily} \cite{altenburger2018monophily} (only some individuals show preferences for labels unrelated to their own); ii) \textit{perception biases} \cite{lee2019homophily} (homophily combined to the minority size of a group are responsible for false consensus or false uniqueness).
Inferring and quantifying individual differences comes as a hard task in complex network analysis.
In presence of strong variations, the mean value of a group is unable to fully describe individual node preferences \cite{cantwell2019mixing}.
Hence, several works focused on multi-scale strategies to estimate homophily.
In \cite{peel2018multiscale}, node similarity is measured as the autocorrelation of a time-series defined as a sequence of node labels visited by a random walker allowing a restart.
The assumption is that random walks can integrate information about paths of all possible lengths, extracting similarities from a higher context than the adjacent neighborhood only.
A similar approach is used in \cite{gutierrez2019multi}, applied in graph classification: multi-hop assortativity is defined as the probability that a randomly selected node and a randomly selected $t$-hop neighbor belong to the same category, where $t$ indicates the time of the visit of the random walker.
In \cite{bassolas2021first}, the focus is on the mean first passage times between preassigned classes of nodes, i.e., the expected number of steps needed for a random walker to visit for the first time a node of a certain class when it starts from a node of another class. This concept is used to estimate nodes' heterogeneity, polarisation and segregation.
%Moving from random walkers, i.e., from the probability to reach nodes, to the real distances between them, \textit{Conformity} \cite{rossetti2021conformity} is a node-centric homophily measure where the multi-scale strategy depends on a distance decay parameter used tuning the similarity score obtained from the neighbors at different hops of distance from a target one.
%The concept of \textit{distance} is also exploited to redefine classic statistical measures - as variance and co-variance -- while trying to describe the relations between the \textit{features of functional aspects of the networks} (i.e., attribute distributions) and the relative underlying network topology \cite{devriendt2020variance}.
%\\ \ \\
\noindent {\bf Dynamics of networks}. In complex networks, when time is involved, differentiating between persistent and instantaneous connections is fundamental.
Friendships relations and scientific collaborations are persistent over time, whereas phone calls and physical proximity interactions can only have a certain duration. Emails, messages on social media or financial transactions are inherently instantaneous.
Hence, several models are needed to properly represent all these different dynamics. The most used representations are the Graph Snapshot (SN), the interval Graph (IG), and the Link Stream (LS).
In SNs, a network is represented as a sequence of graphs, each one analyzed autonomously and independently from the others.
In IGs, the focus is on the characterization of link durations, these ones identified by a start and an end in time.
In LSs, a link is identified only by a pair of nodes and an instantaneous point in time.
Each network representation has its \textit{pros} and \textit{cons}.
SNs can capture significant information but can lead to losses of temporal information, e.g., paths within a snapshot do not respect the time-varying dynamics of interactions. 
IGs allow to define time-respecting paths, where the focus is on the nodes that can be reached from other ones within some observations window; however, choosing the appropriate time window may not be trivial.
More than the others, LSs consider the streams of interactions over time, taking into account both the temporal and the structural nature of interactions.
Recent works tried to further generalize such representations: for instance, the Stream Graph model \cite{latapy2018stream} aims to create a formalism dealing with both instantaneous links and links with duration.
Stream Graphs can easily extend and generalize static centrality measures as the betweenness \cite{simard2021computing}, and analyze empirically the differences in shortest, fastest, foremost time-respecting paths \cite{simard2019computing, simard2021evaluating}.
Recently, augmented stream graphs models have been defining, as in modeling interactions over time with multi-layer structure \cite{parmentier2019introducing}, where the focus is on the definition of new layer centrality measures.

\noindent {\bf Towards temporal homophily estimation}. The temporal dimension of homophily is still little understood in complex network analysis.
Introducing new dimensions in homophily estimation is not trivial.
%it is required to define new homophily conceptual variants and identify new methodological solutions to estimate them in networks.
For instance, a definition of anti-assortative and anti-disassortative mixing patterns is needed while studying homophily in trust or signed networks, where sharing similar values must be related to the positive or negative value of an interaction \cite{rathore2013analyzing}.
Introducing a temporal dimension means dealing with different time representations.
Hence, temporal homophily can present different facets.
Several works addressed different research questions, giving more facets and complexity to the problem.
For instance, some works about degree-degree assortativity describe the evolution patterns of this property.
In \cite{zhou2020universal}, a universal behavior in temporal homophily is explained as follows: \textit{degree assortativity increases at the beginning of network evolution and decreases to a long-lasting stable level}.
This behaviour was observed independently in several domain-specific domains, as in the Bitcoin Transaction Network \cite{kondor2014rich}.
\textit{Temporal homophily} has been defined in \cite{kovanen2013temporal} by leveraging the notion of colored networks. 
Here, homophily is described as the tendency of similar nodes to participate in mixing patterns -- through occurrences of node colors in temporal motifs -- beyond what would be expected from a null model representing the structure of the aggregate network, assuring indeed that dynamic observations are independent of results obtained from a static analysis.
The interplay between structure, attributes, and time has also been addressed in \cite{sepulvado2020predicting} as a relation between a first-order and second-order similarity, where the former one is the common definition of homophily (between attributes sharing similar values), and the latter one is the similarity in trajectories of changes.

\section{$\Delta$-Conformity}
\label{sec:delta_conf}
Our work aims to propose a measure able to characterize homophilic behaviors in presence of time evolving topologies.
During the last decade, several approaches have been proposed to support dynamic network analysis; in this section, we firstly provide a description of the modeling paradigm adopted, then describe the \textit{Conformity} measure and its extension - namely \textit{$\Delta$-conformity} - to time evolving networks. 
%To arrive at our definition of $\Delta$-conformity, we need first to give some context definitions that will be used. Therefore in the subsection \ref{subsec:context}, we present the context definitions for the attribute graph, the stream graph and various path; in the second subsection \ref{subsec:definitionConformity}, we focus on conformity and extension to $\Delta$-conformity.

\subsection{Feature-rich Stream Graphs}
\label{subsec:context}
Two peculiar features characterize the network structures whose homophily we are interested in measuring: (i) their nodes are enriched by categorical attributes and, (ii) their node and edge's sets are allowed to vary as time goes by.
Since such information can be modeled leveraging several different frameworks, it is mandatory to properly define the reference framework used.
For the sake of simplicity, we introduce our model - Feature-rich Stream Graph - through incremental definitions, introducing first node annotations and then temporal dimension. 

When in the presence of a static network whose nodes exposes semantic properties, we define a Node-attributed graph as:
\begin{definition}[\textbf{Node-attributed Graph}]
$\mathcal{G}=(V,E,L)$ is a node-attributed graph, where $V$ is the set of nodes, $E$ the set of edges, and $L$ a set of categorical attributes such that $L(v)$, with $v \in V$, identifies the set of categorical values associated to $v$.
\end{definition}

Stream graph \cite{latapy2018stream} is a powerful modeling formalism that allows us to describe and characterize streams of nodes and edges having an associated lifespan (i.e., appearing and vanishing, even multiple times, during the network life). Therefore, we leverage this mature formalism to integrate the temporal dimension within the Node-attributed graph framework:
%Moreover, since our work explicitly target time evolving network topologies, we conservatively extend such a definition in the realm of the stream graphs:% to capture the both temporal and structural nature of interactions:
\begin{definition}[\textbf{Feature-rich Stream Graph}]
$\mathcal{S}=(T,V,W,E,L)$ is a stream graph, where $T = [\mathrm{A}, \Omega]$ is the set of discrete time instants, with $\mathrm{A}$ and $\Omega$ the initial and final instants, $W \subseteq T \times V$ the set of temporal nodes, $E \subseteq T \times V \times V$ the set of edges such that $(t,uv) \in E$ implies $(t,u) \in W$ and $(t,v) \in W$ and $L$ the set of temporal node attributes such that $L(t,v)$ with $v \in V$ and $t \in T$, identifies the set of categorical values associated to $v$ at time $t$.
\end{definition}

\begin{figure*}[t!]
\centering
{\includegraphics[scale=0.5]{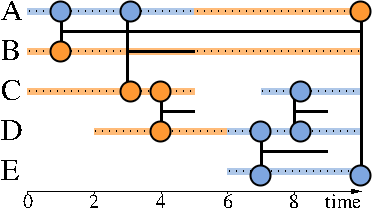}}
  \caption{A Feature-rich Stream Graph with time-varying labels. Vertical and solid horizontal segments identify edges and their duration, respectively; dotted horizontal segments are the node labels, e.g., \textit{A}'s label from $t=0$ to $t=5$ is \textit{blue}, changing in \textit{orange} from that point on.}
    \label{fig:toy_stream}
\end{figure*}

Feature-rich Stream Graphs allow describing those networks whose topology and node-attributes vary as time goes by.
Consider example \ref{fig:toy_stream}, where edges and their duration are identified respectively by vertical and solid horizontal segments, while node-attributes are represented by node colors; a node maintains a label along the dotted horizontal lines until the color changes.
To contextualize the toy example, we can imagine the edge identifying pairwise discussions among users in an online social media, while node-attributes identify individual stances on a given topic (e.g., the left/right leaning of the users involved in a political debate).

\subsection{Multiscale Homophily in Feature-rich Stream Graph}
\label{subsec:definitionConformity}
We describe \textit{$\Delta$-Conformity} as a conservative extension of \textit{Conformity} (previously introduced in \cite{rossetti2021conformity}) to Feature-rich Stream Graph.

Given a Node-Attributed (static) Graph, \textit{Conformity} is a multi-scale homophily measure that accounts for the length of paths connecting node pairs while computing individual degrees of assortativity. % similarity between the attributes of a target node with the other nodes of a network. %; in the process, it leverage a decay parameter to weigh node similarity with the distance among nodes.
To better clarify how \textit{Conformity} - and by extension \textit{$\Delta$-Conformity} - works, we provide its formalization.

\begin{definition}[Conformity]
Given a real number $\alpha$ in $[0, +\infty)$, the \textit{Conformity} score for a node $u\in V$ is defined as:

\begin{equation}
\psi(u,\alpha) = \frac{\sum_{d \in D} \frac{\sum_{v \in N_{u,d}} I_{u,v} f_{v,l_v}}{\vert N_{u,d} \vert d^\alpha}}{\sum_{d \in D} d^{-\alpha}},
\label{eq:conformity} 
\end{equation}

where:
\begin{itemize}
    \item $D$ is the maximum distance among all node pairs $(u,v) \in V$; % $max(\{dist(i,j)|i,j \in V\})$
    \item $N_{u,d}$ is the set of $u$'s neighboring nodes at distance $d$, i.e., the nodes that can be reach after $d$ number of hops from the target node $u$;
    \item$\alpha$ tunes the relevance of nodes at distance $d$ by the source $u$;
    \item $I_{u,v}$ is the indicator function comparing the attribute values of a node $u$ and a node $v$
        \begin{equation}
                I_{u,v} = 
            \left\{
                \begin{array}{ll}
                    1  & \mbox{if } l_u=l_v \\
                    -1 & \mbox{otherwise},
                \end{array}
            \right.
            \label{eq:Iuv}
        \end{equation}
    \item $f_{u,l_u}$ is the similarity function computing the ratio of $u$'s first-order neighbors (namely $\Gamma(u)$) sharing its same attribute value $l_u$ - a ratio forced to the range in $(0, 1]$ by imposing its value to 1 when the numerator nullifies
        \begin{equation}
            f_{u,l_u} = \frac{ \vert \{v\vert v \in \Gamma(u) \land l_u=l_v\}\vert}{\vert\Gamma(u)\vert}.
        \end{equation}
    In the measure we compute the similarity function on each $u$'s neighbor $v$, namely $f_{v,l_v}$.
\end{itemize}

The final score is thus normalized to ensure that \textit{Conformity} lies in the range $[-1, 1]$.

\end{definition}

In its original definition \cite{rossetti2021conformity}, the distance function adopted by \textit{Conformity} is the one computing network geodesic paths.
To cope with temporal constraints introduced by dynamically evolving topologies, such a choice needs to be revised.

While adopting the Feature-rich Stream Graph model, we are particularly interested in observing interactions occurring at least once every $\Delta$ units of time -- where $\Delta$ is a time interval of a certain duration in $T$.
Therefore we define \textit{$\Delta$-Conformity} as a $\Delta$-analysis \cite{latapy2018stream} approach:

\begin{definition}[\textbf{$\Delta$-analysis: Stream Graph}]
$\mathcal{S}_\Delta=(T_\Delta,V,W_\Delta,E_\Delta, L_\Delta)$ is the attributed stream graph such that $T_\Delta = [\mathrm{A} + \Delta, \Omega - \Delta]$, an edge $(t', uv) \in E_\Delta$, with $u \in W_\Delta$ and $v \in W_\Delta$, is present at a time $t'$ in $\mathcal{S}_\Delta$ and each node $v \in V$ has a set of categorical values $L(t',v)$ whenever it is present in $\mathcal{S}$ at a time $t$ in $[t' + \Delta]$ \footnote{The interval considered in \cite{latapy2018stream} is $[t' - \frac{\Delta}{2}, t' + \frac{\Delta}{2}]$; in our formalism, we remain consistent with the analyses performed in the experimental section.}.

\end{definition}

Once fixed such a reference framework we can define \textit{$\Delta$-Conformity}:
\begin{definition}[$\Delta$-Conformity]
Given a real number $\alpha$ in $[0, +\infty)$, a temporal id $t\in T$, and a time window $\Delta$, the \textit{$\Delta$-Conformity} score for a node $u\in V$ is defined as the value of Conformity($\alpha$, u) where:
\begin{itemize}
    \item[-] the Feature-rich Stream Graph is restricted to $\mathcal{S}_\Delta=(T_\Delta,V,W_\Delta,E_\Delta, L_\Delta)$ with $T_\Delta = [t, t + \Delta]$;
    \item[-] the distance among $u,v \in V$ is the length of a time-respecting path connecting the two (if it exists, $\infty$ otherwise). 
\end{itemize} 
\end{definition}

The proposed definition introduces two peculiarities that make \textit{$\Delta$-Conformity} more complete w.r.t. its predecessor: (i) the concept of \emph{memory} (modeled with $\Delta$), and (ii) of \emph{preferential} temporal interaction patterns (made explicit by the distance function). 
The former aspect defines the temporal bounds for computing homophily: the underlying idea is that the communication chains used to evaluate nodes' similarity need to be bound to a specific temporal window (namely, as time pass new interactions in the chain lose their relevance for the source node).
The latter aspect instead focuses on how such chains are built.
If in static networks the simplest function to compute distances among two nodes is computing the length of the shortest path connecting them, several alternative definitions can be used to reach a similar goal in dynamic networks.

Indeed, the notion of \textit{distance} is crucial while emphasizing the dynamic nature of graphs.
Paths in stream graphs have both a length and a duration.
A time-respecting path can be defined as in the following:

\begin{figure*}[t!]
\centering
\subfloat[Shortest paths]{\includegraphics[scale=0.5]{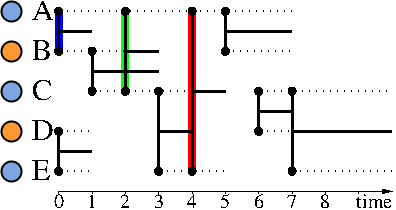}} \qquad
\subfloat[Foremost paths]{\includegraphics[scale=0.5]{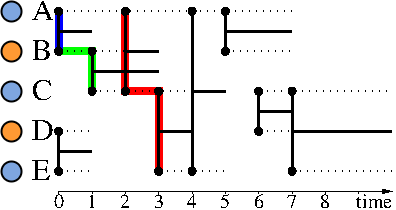}}
  \caption{\textit{Time respecting paths.} A comparison of shortest (a) and foremost (b) paths starting from the node A, at $t=0$, and targeting nodes reachable within a time window  $\Delta=5$. Assuming the set $blue=(A,C,E)$ and $orange=(B,D)$ of nodes sharing a same label and applying the proposed measure we observe that: (a) shortest paths identify an assortative behavior $\Delta-Conformity(A)=0.33$ while, (b) foremost ones identify a disassortative one, $\Delta-Conformity(A)=-0.41$.}
    \label{fig:toy2}
\end{figure*}

\begin{definition}[\textbf{Time-respecting path}]
In a stream graph $S=(T,V,W,E)$, a sequence $(t_0, u_0, v_0),\dots, (t_k, u_k, v_k)$ of elements $T \times V \times V$, such that $u_0 = u$, $v_k = v$, $t_o \geq \alpha$, $t_k \leq \omega$, for all $i, t_i \leq t_{i+1}$, $v_i = u_{i+1}$, and $(t_i, u_i, v_i) \in E$, $[\alpha, t_0] \times {u} \subseteq W$, $[t_k, \omega] \times {v} \subseteq W$, and for all $i, [t_i, t_{i+1}] \times {v_i} \subseteq W$ 
is a time-respecting path from $(t_0, u_0) \in W$ to $(t_k, v_k) \in W$ of length $k$ and duration $t_k - t_0$.
\end{definition}

This leads to different notions for the cost of reaching nodes, distinguishing between different notable typologies of time-respecting paths. 
As an example, considering a Stream Graph $S=(T,V,W,E)$ we can focus on:
\begin{itemize}
    \item[-] \emph{Shortest path}: $\mathcal{P}$ is a shortest path from $(t_0, u_0) \in W$ to $(t_k, v_k) \in W$  if it has minimal length $k$;
    \item[-] \emph{Fastest path}: $\mathcal{P}$ is a fastest path from $(t_0, u_0) \in W$ to $(t_k, v_k) \in W$  if it has minimal duration $t_k - t_0$;
    \item[-] \emph{Foremost path}: $\mathcal{P}$ is the path from $(t_0, u_0) \in W$ to $(t_k, v_k) \in W$  that, independently from its length and duration, allows to reach first the destination.
\end{itemize}

While used within the framework of \textit{$\Delta$-Conformity}, those three path types - that represent only a subset of interesting ones - can profoundly affect the observed homophily/heterophily of individual nodes.

Consider for instance the toy example reported in \ref{fig:toy2} involving the temporal interactions among five nodes $V = [A,B,C,D,E]$ in $T=[0,10]$. 
There we assume $\Delta=5$ and compare shortest and foremost paths starting from $A$, namely:
\begin{itemize}
    \item[-] Shortest paths: 
    \begin{itemize}
        \item[] \textbf{(A,B)} = [(A,B,0)]
        \item[] \textbf{(A,C)} = [(A,C,2)]
        \item[] \textbf{(A,E)} = [(A,E,4)]
    \end{itemize} 
    
    \item[-] Foremost paths:
    \begin{itemize}
        \item[] \textbf{(A,B)} = [(A,B,0)]
        \item[] \textbf{(A,C)} = [(A,B,0),(B,C,1)]
        \item[] \textbf{(A,E)} = [(A,C,2),(C,E,3)]
    \end{itemize}
\end{itemize}
Considering two possible node labels, blue and orange, assigned respectively to the node sets $(A,C,E)$ and $(B,D)$ we can observe that \textit{$\Delta$-Conformity} ($\alpha=1$) unveils completely different homophilic patterns for node $A$: assortative while considering shortest paths, disassortative for foremost paths.

Such a counterintuitive behavior is tied to the different distances that the two time-respecting paths generate, and its interpretation can be easily clarified, providing more semantic to our example.
Let us assume that the previous stream-graph models a word of mouth-like diffusion phenomenon and that users' labels describe opposing stances regarding the discussed topic: every time the content is passed to a peer having a different opinion from the source, its original value is reduced.
Foremost paths aim to reach all the available users as early as possible, disregarding that the original content might reach same-stance users only after being dismantled piece after piece.  
On the other hand, shortest paths aim to minimize such an effect by cutting down the length of message passing chains.

Indeed, one of the advantages of $\Delta$-analysis lies in the opportunity of analyzing the temporal trends of a given indicator.
For \textit{$\Delta$-Conformity}, this means observing -- node-wise -- how homophilic/heterophilic behaviors unfold as time goes by.
Leveraging such rationale, in the following section we underline how peculiar \textit{$\Delta$-Conformity} trends can characterize different actors of real-world evolving complex systems.

\section{Experiments}
\label{sec:exp}

In this section, we explore \textit{$\Delta$-Conformity} properties by applying the measure to several real-world datasets from different domains.
We focus on finding heterogeneous mixing behavior by studying the characteristics of single nodes or classes of nodes over time.
We also focus on testing the statistical significance of the emerging trends, i.e., whether they are interesting patterns of the datasets or whether distributions from randomized networks let similar patterns emerge.
%This problem is similar to the ones occurring in community detection, where also from random graphs there can be reached high values of modularity, for instance \cite{fortunato2016community}.
In particular, this quantitative analysis is performed on the Bitcoin Network (cf. \textit{Bitcoin Network}).
%Conversely, we will comment the results from Copenhagen Network Study and the SocioPattern datasets only referring to the original data papers for their interpretations. 

\begin{figure*}[t!]
\centering
\subfloat[$\Delta=1$]{\includegraphics[scale=0.43]{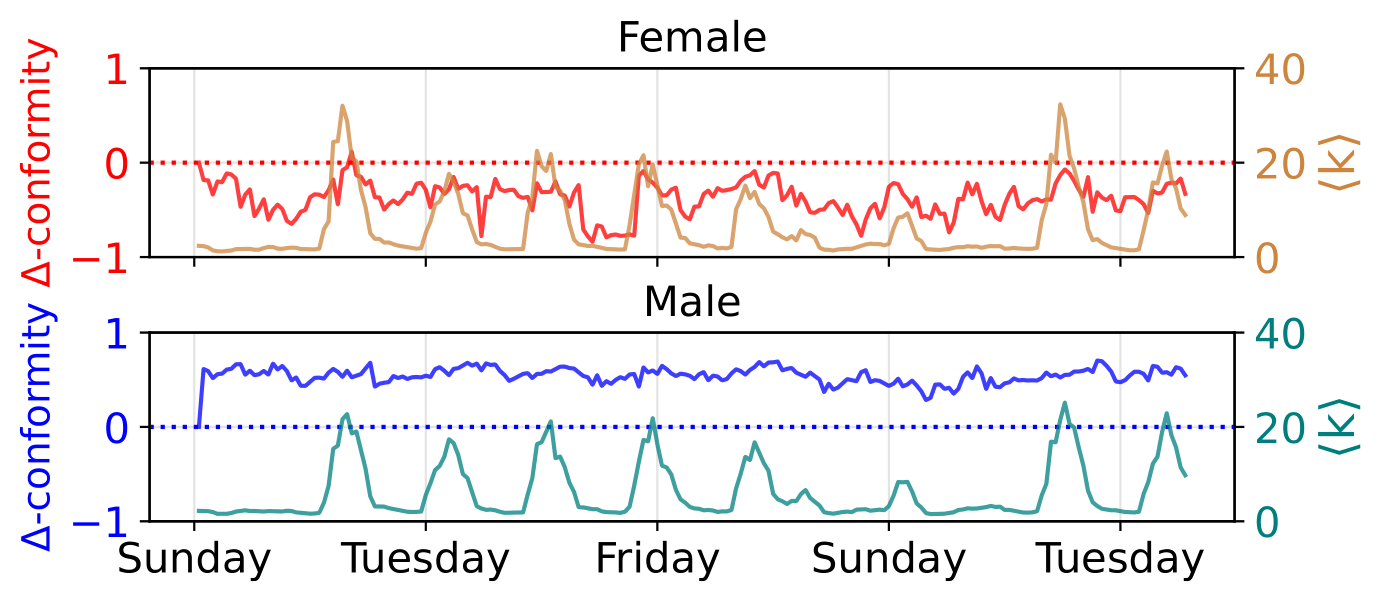}}
\subfloat[$\Delta=8$]{\includegraphics[scale=0.43]{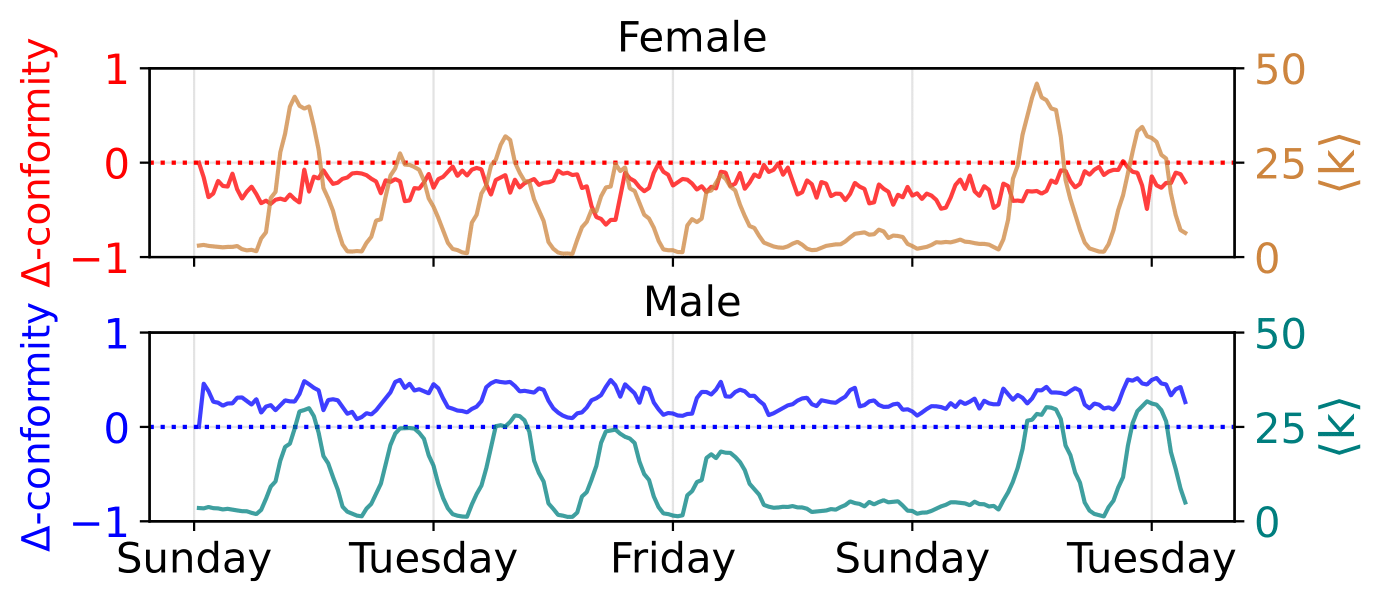}}
  \caption{\textit{Copenhagen Network Study}. \textit{$\Delta$-Conformity} by gender using (a) $\Delta$=1 and (b) $\Delta$=8; the two y-axes indicate the average \textit{$\Delta$-Conformity} score of the group (left axis) and the average degree $\left\langle k \right\rangle$ of the group (right axis) of the category. The horizontal colored lines, i.e. \textit{$\Delta$-Conformity}=0, highlight the uniformly mixed behaviour.}
    \label{fig:copen}
\end{figure*}

In all the experiments we focus on the \textit{shortest paths} only, fixing the decay parameter $\alpha$ to $2$ \cite{rossetti2021conformity}. Different values of $\Delta$ are used with respect to the analysis purposes or the dataset nature.
%\\ \ \\
\noindent {\bf Copenhagen Network Study}. We consider the 700 (male and female) university students participating in the Copenhagen Network Study \cite{sapiezynski2019interaction}, whose social interactions are estimated via Bluetooth signal strength.
In \cite{rossetti2021conformity}, \textit{Conformity} by gender was measured in a daily-aggregated static network, where males were on average more assortative than females.
We aim to provide more insights into students' daily routines, e.g., whether mixing patterns differ by day and night or by weekdays and weekends.
Hence, we consider a dynamic network covering little more than one week, i.e., from Sunday 12 am to Tuesday 2 pm.
Timescale aggregation window is by \textit{hours}, then we consider two different $\Delta$, 1 and 8, where interactions and paths occur within a window of 1 or 8 hour(s).
Fig. \ref{fig:copen} reports the average \textit{$\Delta$-Conformity} trends of male and female students.
Similar to the static aggregated analysis \cite{rossetti2021conformity}, females present a disassortative trend while male trends are assortative.
Interestingly, the average values differ when using two different values of $\Delta$.
Using larger windows implies including more links in the analysis, i.e., higher average degree $\left\langle k \right\rangle$.
This can explain why the average \textit{$\Delta$-Conformity} scores are closer to the horizontal lines: the higher the aggregation of links, the lesser the distinctiveness of groups' mixing patterns, where \textit{$\Delta$-Conformity}=0 corresponds to the uniformly mixed behavior.
However, larger windows let emerge periodicity. In particular, the trend of male students correlates with the average degree trend of the group with $\Delta$=8.
Thus, higher values of $\Delta$ can suggest a broad point of view that can capture the circadian nature of people interactions, and both the two network indicators, i.e., \textit{$\Delta$-Conformity} and average degree $\left\langle k \right\rangle$, can help to describe these behaviors better.

\begin{figure*}[t!]
\centering
\subfloat[Hospital ward]{\includegraphics[scale=0.39]{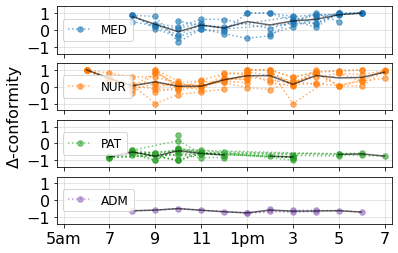}}
\subfloat[Primary school]{\includegraphics[scale=0.39]{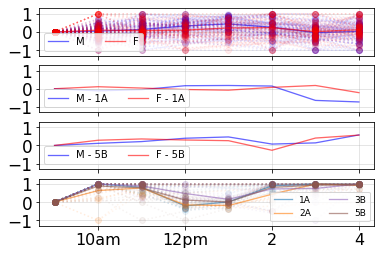}}
\subfloat[High school]{\includegraphics[scale=0.39]{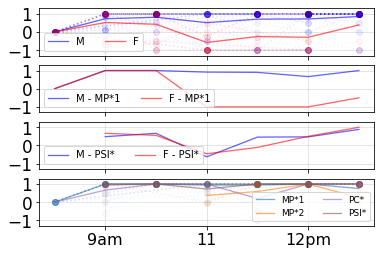}}
  \caption{\textit{SocioPatterns}. (a) \textit{$\Delta$-Conformity} by hospital category; (b-c) from top down: \textit{$\Delta$-Conformity} by gender: trends showing the whole network (first subplot) and the subgraphs of two distinct classes (second, third subplots); \textit{$\Delta$-Conformity} by class (last subplot); (a) black or (b-c) colored lines represent the average value of the groups while points are the values of individual nodes.} 
    \label{fig:sociopat}
\end{figure*}

\noindent {\bf SocioPatterns} \footnote{http://www.sociopatterns.org/}. We consider three datasets from the SocioPatterns collection: \textit{Hospital ward} \cite{vanhems2013estimating} is a set of contacts between patients and health-care workers; \textit{Primary school} \cite{stehle2011high} and \textit{High school} \cite{mastrandrea2015contact} are two sets of contacts and friendship relations between children/high-school students.
We focused on one day in all three networks. Timescale aggregation window is by \textit{hours}. We set $\Delta$=1 to capture fine-grained activities, e.g., lunch breaks.
Fig. \ref{fig:sociopat} sums up the analysis on these networks.

The first panel -- Fig. \ref{fig:sociopat} (a) -- introduces the hospital ward dataset.
Black lines represent the average \textit{$\Delta$-Conformity} trends of the groups, while colored points are individual nodes, and the corresponding dotted colored lines follow each node for tracking the evolution of the score (medical doctors in blue, nurses and nurse' aids in orange, patients in green and administrative staff in purple).
The analysis starts from $6am+\Delta$, when contacts are between nurses and nurse' aids only; this results in perfectly assortative group behavior.
When nurses and nurse' aids start to visit patients \cite{vanhems2013estimating} ($7am+\Delta$), nodes' mixing splits into two branches, one exhibiting assortativity and the other one disassortativity.
Hence, individual differences become visible, and the average \textit{$\Delta$-Conformity} score of the group, i.e., the black line, is no longer useful for capturing group heterogeneity.
On average, patients remain disassortative all day; since \textit{all rooms but 2 were single-bed rooms} \cite{vanhems2013estimating}, also the assortative behaviour of some patients at $10am+\Delta$ is explained.
The absence of patients at certain times of the day can be motivated with the lunch break (e.g., $1 pm+\Delta$) or with a pause of the visit time (e.g., $4 pm+\Delta$).
The few administrative staff members are disassortative all the time.

The other two panels -- Fig. \ref{fig:sociopat} (b-c) -- introduce the analysis on the two schools.
We measure homophily by gender and homophily by class.
From top down, the first three subplots (of both schools) focus on homophily by gender.
The first subplot focuses on the whole network visualization, while the other two focus on two selected classes' subgraphs.
Colored solid lines represent the average trends of the groups, while points are the scores of individual nodes.
Interestingly, primary and high school students behave differently with respect to the gender attribute.
Primary students are uniformly mixed. Conversely, the behavior of high school students is similar to what we have already observed in the Copenhagen college network: males are on average more assortative than females.
Focusing on a subset of classes (second, third subplots) allows us to explain in which classes this divergent behavior among groups is stronger; e.g., mathematics and physics students (\textit{MP*1}) \cite{mastrandrea2015contact} present more differences than engineering students (\textit{PSI*}).
The last subplot of each school shows \textit{$\Delta$-Conformity} by class.
As expected, high schools students are strongly assortative with respect to the class, motivated by the fact that students of different classes do not know each other.
A more interesting pattern occurs in primary school.
In line with the data described in the reference paper \cite{stehle2011high}, the disassortative trends occurring from $12 am+\Delta$ to $1 pm+\Delta$ can be explained through the children spatio-temporal trajectories: children move from their rooms to the playground or cafeteria for lunch; hence it is more likely for a child to enter in contact with children of another class.

\begin{figure*}[t!]
\centering
\subfloat[]{\includegraphics[scale=0.36]{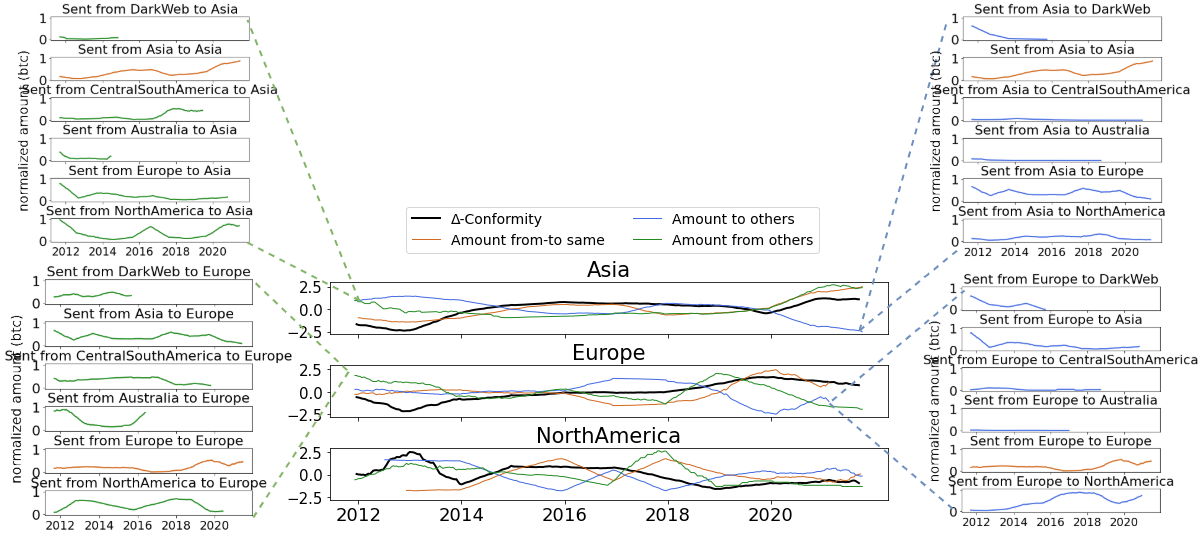}}
\quad
\subfloat[]{\includegraphics[scale=0.35]{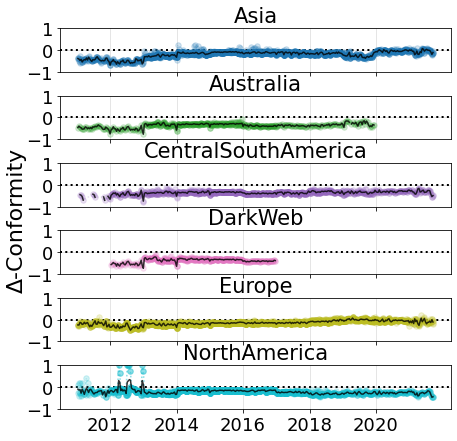}}
\subfloat[]{\includegraphics[scale=0.31]{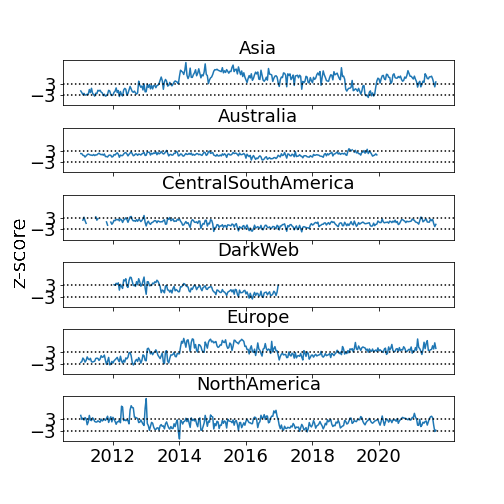}}
  \caption{\textit{Bitcoin Transaction Network - Location Attribute}. (a) Comparison between the \textit{$\Delta$-Conformity} and the Bitcoin amount trends, with a focus on the amounts sent to/from Asia and Europe from/to other classes; (b) \textit{$\Delta$-Conformity} of different classes w.r.t. to the location attribute, and (c) their z-score trends. } 
    \label{fig:bitcoin_location}
\end{figure*}
\begin{figure*}[t!]
\centering
\subfloat[]{\includegraphics[scale=0.36]{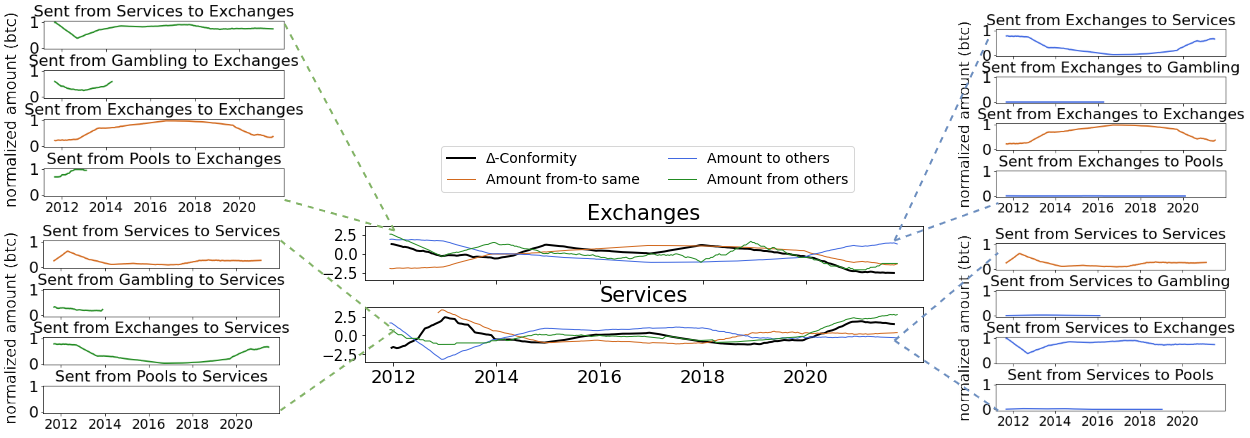}}
\quad
\subfloat[]{\includegraphics[scale=0.35]{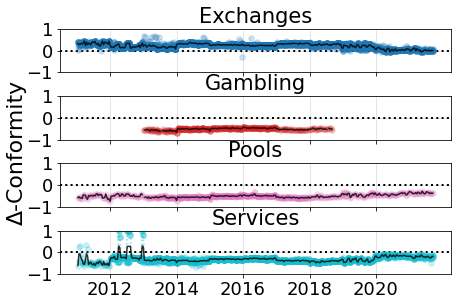}}
\subfloat[]{\includegraphics[scale=0.31]{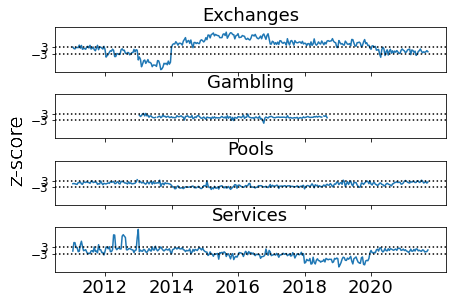}}
  \caption{\textit{Bitcoin Transaction Network - Category Attribute}. (a) Comparison between the \textit{$\Delta$-Conformity} and the Bitcoin amount trends, with a focus on the amounts sent to/from Exchanges and Services from/to other classes; (b) \textit{$\Delta$-Conformity} of different classes w.r.t. to the category attribute, and (c) their z-score trends.} 
    \label{fig:bitcoin_category}
\end{figure*}
%\\ \\ \\
\noindent {\bf Bitcoin Network}. We consider the network of Bitcoin transactions extracted from the blockchain between blocks 0 and 667542 (January 2021). Actors (groups of addresses) are identified using the standard co-input heuristic \cite{harrigan2016unreasonable,remy2017tracking}, and we create a daily aggregated network in which there is an edge between two actors if there is an observed transaction between them on that day. We filter the network to keep only the top 100 actors with the most transactions.
Similarly to \cite{jourdan2018characterizing}, we label each user with its \textit{category} from the WalletExplorer\footnote{https://www.walletexplorer.com/}, among Exchanges, Gambling, Pools, and Services.
Moreover, we label each node with an attribute identifying actor's headquarter \textit{location}, among Asia, Australia, Central \& South America, Europe, North America, and those illegal websites without a physical location labeled as \textit{Dark Web}.
Each link is weighed with the amount of Bitcoin sent to/from each node.
Assuming an undirected network while estimating node \textit{$\Delta$-Conformity}, we distinguish between the amount sent to/from each group only for adding more insights in the interpretation of \textit{$\Delta$-Conformity} trends.
Remember that \textit{Conformity} does not consider link weights in its measurement.

We study the two attributes separately.
We consider $\Delta$=1, hence links occurring day by day.
Fig. \ref{fig:bitcoin_location} and Fig. \ref{fig:bitcoin_category} sum up the analysis on \textit{$\Delta$-Conformity} by location and by category, respectively.
In both figures, \textit{$\Delta$-Conformity} trends are plotted together with the amount of Bitcoin sent to and from each location and category, respectively -- Fig. \ref{fig:bitcoin_location} (a), Fig. \ref{fig:bitcoin_category} (a).
Groups' amounts are normalized w.r.t. the total amount of money sent in each $\Delta$ window considered.
This information is shown in detail in the subplots on the left and right sides of the panel (left, amount sent from the other groups to the target one; right, amount sent from the target group to the other ones).
For instance, Asian actors tend to send less money to other locations and exchange bigger quantities between other Asian actors; similarly, European actors tend to send more money among themselves, but other locations send less money to them.
\textit{$\Delta$-Conformity} trends seem to follow the same pattern, e.g., more assortative when the amount sent from-to the same attribute value are bigger.
Similarly, Exchanges actors mainly play among themselves, and their average \textit{$\Delta$-Conformity} trend is assortative on average, while Services actors mainly play with Exchanges actors, and their average \textit{$\Delta$-Conformity} trend is disassortative.

While aligning \textit{$\Delta$-Conformity} and Bitcoin trends in one plot, we lose the original scale of \textit{$\Delta$-Conformity} scores.
Hence, \textit{$\Delta$-Conformity} scores are also plotted in Fig. \ref{fig:bitcoin_location} (b) and Fig. \ref{fig:bitcoin_category} (b), lying in the original range of values.
Locations are slightly disassortative or uniformly mixed over time -- Fig. \ref{fig:bitcoin_location} (b).
Categories are mainly disassortative, except for Exchange, that are on average slightly assortative -- Fig. \ref{fig:bitcoin_category} (b).

A value of \textit{$\Delta$-Conformity} below zero indicates that actors of the corresponding category do not interact mostly with actors of the same category, i.e., they do not form a \textit{closed club}. However, they might still be interacting more frequently with actors of the same category than expected at random. To check this hypothesis, we use a bootstrap approach and z-scores to check if the \textit{$\Delta$-Conformity} is nevertheless significantly larger than expected by chance, thus pointing towards a preference for actors of the same category.
Hence, panel (c) of both figures shows the z-scores trends assessing the reliability of \textit{$\Delta$-Conformity} trends.
Z-scores are obtained by comparing point by point the \textit{$\Delta$-Conformity} score of the original graph to a null distribution of 200 rewired networks using a configuration model \cite{molloy2011critical, gauvin2018randomized}; 
We use the following formula for the comparison:

\begin{center}
    $z= \frac{x - \mu}{\frac{\sigma}{\sqrt n}}$
\end{center}

where $x$ is the \textit{$\Delta$-Conformity} value in the original graph, $\mu$ is the average \textit{$\Delta$-Conformity} value from the ensemble of rewired networks, $\sigma$ is the standard deviation of the ensemble, and $n$ is the number of nodes having $l$ as label, e.g., having Asia as location.
Horizontal lines at values 3 and -3 are supporting to determine the statistical significance of original \textit{$\Delta$-Conformity} scores.

We can suppose that the expected behaviour of a rare/less frequent category is disassortative, because few of the encountered nodes are of the same category, whatever the distance.
For instance, the very few \textit{Australian} or \textit{DarkWeb} nodes are highly disassortative, but this behaviour is not significant.
Conversely, the \textit{Asian} nodes are still disassortative -- they interact to each other and to other 5 classes as well -- but their z-scores above 3 are meaning that this behaviour is still higher than expected -- cf. Fig. \ref{fig:bitcoin_location} (c).
Similarly, Exchanges and Services \textit{$\Delta$-Conformity} trends are more statistically reliable than Gambling and Pools -- Fig. \ref{fig:bitcoin_category} (c).

\section{Conclusions}
\label{sec:conc}

In this work we explored node-centric mixing pattern estimation in feature-rich stream graphs by extending the recent multi-scale, path-aware measure of \textit{Conformity} \cite{rossetti2021conformity}.
To lift the unrealistic assumption of a fixed network topology, we leveraged the formalism of the $\Delta$-analysis of stream graphs \cite{latapy2018stream}.
Re-framing \textit{Conformity} in such a framework allowed us to propose a more fine-grained and dynamic node-wise homophily estimator, i.e., \textit{$\Delta$-Conformity}, able to cope with time-varying paths, among shortest, fastest, foremost, etc., and with both static and varying node categorical attributes.

While analyzing \textit{$\Delta$-Conformity} trends in several social interaction networks, we found that the mixing behaviour of node can change over time. Such changes coincide with contextually reasonable everyday patterns, from the working hours of medical staff to the lunch break of primary school children.
Moreover, a broad case-study on the Bitcoin Transaction Networks convinced us that \textit{$\Delta$-Conformity} could be applied on several and different domains, not only social networks.

As future works, we plan to apply \textit{$\Delta$-Conformity} for hot topics in computational social science, as in multi-scale echo-chamber identification \cite{morini2021toward}, or for monitoring users in sensible online communities, e.g., mental-health related ones \cite{joseph2021cognitive}. We expect that the same mechanisms applied on Bitcoin network could allow us to identify cliques of actors working together to manipulate the market or the formation of national markets, for instance in a country adopting Bitcoin as local tender such as El Salvador.

\section*{Acknowledgments}

This work is supported by the European Union – Horizon 2020 Program under the scheme "INFRAIA-01-2018-2019 – Integrating Activities for Advanced Communities", Grant Agreement n.871042, "SoBigData++: European Integrated Infrastructure for Social Mining and Big Data Analytics" (\url{http://www.sobigdata.eu}) and by the CHIST-ERA grant CHIST-ERA-19-XAI-010, by MUR (grant No. not yet available), FWF (grant No. I 5205), EPSRC (grant No. EP/V055712/1), NCN (grant No. 2020/02/Y/ST6/00064), ETAg (grant No. SLTAT21096), BNSF (grant No. \begin{otherlanguage*}{russian}КП-06-ДОО2/5\end{otherlanguage*}).

This work is partially supported by BITUNAM Project ANR-18-CE23-0004.

\nolinenumbers

%This is where your bibliography is generated. Make sure that your .bib file is actually called library.bib
\bibliography{library}
\bibliographystyle{plain}

\end{document}